\documentstyle[12pt]{article}

\newcommand{\be}{\begin{equation}}  
\newcommand{\ee}{\end{equation}}
\newcommand{\ba}{\begin{eqnarray}}  
\newcommand{\ea}{\end{eqnarray}}

\begin{document}
\vskip 0.2 cm
\centerline{{\bf{Quenched Hadron Spectrum and Decay Constants}}}
\centerline{{\bf{on the Lattice}}}
\centerline{\it L.~Giusti\footnote{e-mail~:~giusti@sabsns.sns.it}}
\centerline{\it Scuola Normale Superiore, P.zza dei Cavalieri 7 and}
\centerline{\it INFN, Sezione di Pisa, 56100 Pisa, Italy.}

\abstract{In this talk we present the results obtained from a study of 
${\cal O}(2000)$ (quenched) lattice configurations from
the APE collaboration, at $6.0\le\beta\le 6.4$, using both the Wilson 
and the SW-Clover fermion action. We determine the light
hadronic spectrum and the meson decay constants. For the light-light systems we 
find an agreement with the experimental data of $\sim 5\%$ for mesonic masses 
and $\sim 10\%-15\%$ for baryonic 
masses and pseudoscalar decay constants; a larger deviation is present 
for the vector decay constants. For the heavy-light decay constants we find
$f_{D_s}=237 \pm 16 $~MeV, $f_{D} = 221 \pm 17 $~MeV ($f_{D_s}/f_D=1.07(4)$), 
$f_{B_s} = 205 \pm 35 $~MeV,
$f_{B} = 180 \pm 32 $~MeV ($f_{B_s}/f_B=1.14(8)$),
in good agreement with previous estimates.}

\vspace{3mm}

\noindent
{\bf Introduction}   \\

\noindent
The aim of this exposition is to describe the recent results 
obtained from the APE collaboration on the hadronic
spectrum and the meson decay constants.
A full description of the method used and the results obtained are reported
in \cite{mio1,mio2}.\\ 
The lattice technique has proved a very effective theoretical tool to determine
phenomenological quantities such as the mass spectrum and weak decay matrix
elements. Unlike other approaches, it does not (in principle) suffer from
uncontrolled approximations. However, in practice, one is forced to work on a
lattice of (i) finite size, with (ii) a finite lattice spacing, and (iii)
unphysically large masses for the light quarks. 
Also the quenched approximation is often used in lattice studies.\\
The aim of the high statistics simulations \cite{mio1,mio2} is to study the 
main systematic errors present in the extraction of the light 
hadronic spectrum and  meson decay constants. It is important to study  the
systematics due to (i), (ii) and (iii) before the effects of the quenched 
approximation can be correctly understood.\\
The results reported in this talk are obtained from ${\cal O}(2000)$ (quenched)
lattice configurations from the APE collaboration, for different lattice
volumes at $6.0 \le \beta \le 6.4$ using both the Wilson action and the  
SW-Clover fermion action. The main parameters used in each simulation are 
listed in table~1.         
The values of beta have been chosen:
\par a) small enough to obtain 
accurate results on  reasonably large physical volumes;
\par b) large enough to be in the scaling region.\\ 
In this  range of $\beta$ we cannot draw any conclusion about $a$ dependence
of hadron masses and pseudoscalar decay constants. On the other hand, we find
that the quenched approximation gives a resonable agreement with the
experimental values, when the comparison is possible.\\
The main physical results of our study have been given in the abstract.
\\  

\vspace{1mm}

\noindent
{\bf Lattice Details}  \\

\noindent
\setlength{\tabcolsep}{.18pc}
\begin{table}
\label{tab:uno}
\begin{center} 
\caption{Summary of the parameters of the runs analyzed.}
\begin{tabular}{||c|ccccccccc||}
\hline\hline         
&C60b&C60a&W60&C62a&W62a&C62b&W62b&W64&C64\\
\hline
$\beta$&$6.0$&$6.0$&$6.0$&$6.2$ &$6.2$&$6.2$&$6.2$&$6.4$&$6.4$\\
Action & SW & SW & Wil & SW & Wil& SW & Wil & Wil & SW \\
\# Confs&170&200&320&250&250&200&110&400&400\\
Volume&$18^3\times 64$&$18^3\times 64$&$18^3\times 64$&$24^3\times 64$& $24^3\times 64$ 
&$18^3\times 64$&$24^3\times 64$&$24^3\times 64$&$24^3\times 64$\\
\hline
$K_l$&  -   &  -   &  -   &0.14144&0.1510&    -   &  -   &0.1488&0.1400\\
     &0.1425&0.1425&0.1530&0.14184&0.1515& 0.14144&0.1510&0.1492&0.1403\\
     &0.1432&0.1432&0.1540&0.14224&0.1520& 0.14190&0.1520&0.1496&0.1406\\
     &0.1440&0.1440&0.1550&0.14264&0.1526& 0.14244&0.1526&0.1500&0.1409\\
\hline
$k_H$&0.1150&  -   &0.1255&0.1210 &0.1300&   -    &0.1300&  -   &   -  \\    
     &0.1200&  -   &0.1320&0.1250 &0.1350&   -    &0.1350&  -   &   -  \\    
     &0.1250&  -   &0.1385&0.1290 &0.1400&   -    &0.1400&  -   &   -  \\    
     &0.1330&  -   &0.1420&0.1330 &0.1450&   -    &0.1450&  -   &   -  \\    
     &  -   &  -   &0.1455&   -   &   -  &   -    &0.1500&  -   &   -  \\     
\hline
& \multicolumn{8}{c}{Light-light mesons with zero momentum}& \\
$t_1 - t_2$ & 15-28 & 15-28 & 15-28 & 18-28 & 18-28 & 18-28 & 18-28 & 24-30 & 24-30 \\
\hline
& \multicolumn{8}{c}{Heavy-light mesons with zero momentum}& \\
$t_1 - t_2$ & 15-28 & 15-28 & 15-28 & 20-28 & 20-28 &   -   & 20-28 &  - &  - \\
\hline
& \multicolumn{8}{c}{Baryons with zero momentum}& \\
$t_1 - t_2$ & - & 12-21 & 12-21 & 18-28 & 18-28 & 18-28 & 18-28 & 22-28 & 22-28 \\
\hline
$a^{-1}_{K^*}$& 2.00(10) & 1.98(8) & 2.26(5) & 2.7(1) & 3.00(9) & 3.0(3) & 3.0(1) & 
4.1(2) & 4.0(2) \\  
\hline
\hline
\end{tabular}
\end{center}
\end{table}
Hadron masses and decay constants have been extracted from two-point
correlation functions in the standard way. For the meson masses and decay 
constants we have computed the following propagators:   
\begin{eqnarray}\label{scalare}
G_{55}(t) & = & \sum_{x}\langle P_5(x,t)P_5^\dagger(0,0) \rangle\; ,\nonumber\\
G_{05}(t) & = & \sum_{x}\langle A_0(x,t)P_5^\dagger(0,0) \rangle\; ,
\end{eqnarray}
where
\begin{eqnarray}
P_5(x,t)    & = & i\bar{q}(x,t)\gamma_5q(x,t)\; ,\nonumber\\
A_\mu(x,t)  & = & \bar{q}(x,t)\gamma_\mu\gamma_5q(x,t)\; .\nonumber 
\end{eqnarray}   
and the following propagators of the vector states:
\begin{equation}\label{vettore}
G_{ii}(t) = \sum_{i=1,3}\sum_{x}\langle V_i(x,t)V_i^\dagger(0,0) 
\rangle\; ,
\end{equation}
where 
\[
V_i(x)   =  \bar{q}(x,t)\gamma_i q(x,t)\; .
\]
In order to determine the baryon masses we have evaluated the following 
propagators:
\begin{eqnarray}\label{barioni}
G_n(t) & = & \sum_{x}\langle N(x,t)N^{\dagger}(0,0) \rangle
\; ,\nonumber\\
G_{\delta}(t) & = & \sum_{x}\langle \Delta_\mu(x,t)
\Delta_\mu^{\dagger}(0,0) \rangle\; ,
\end{eqnarray}
where 
\begin{eqnarray}
N              & = &\epsilon_{abc}(u^aC\gamma_5 d^b)u^c\nonumber\\
\Delta_\mu& = & \epsilon_{abc}(u^aC\gamma_\mu u^b)u^c \; .\nonumber
\end{eqnarray}
We have fitted the zero-momentum correlation functions in eqs.~\ref{scalare}, 
\ref{vettore} and \ref{barioni} to a 
single particle propagator with $cosh$ or $sinh$ in the case of mesonic
and axial-pseudoscalar correlation functions
and with an $exp$ function in the case of the baryonic correlation functions
\begin{eqnarray}
G_{55}(t) & = & \frac{Z^{55}}{M_{PS}}\exp(-\frac{1}{2}M_{PS}T)
\cosh(M_{PS}(\frac{T}{2}-t))\; ,\nonumber\\
& & \nonumber\\
G_{ii}(t) & = & \frac{Z^{ii}}{M_V} \exp(-\frac{1}{2}M_{V}T)
\cosh(M_{V}(\frac{T}{2}-t))\; ,\nonumber\\
& & \label{funzfit}\\
G_{05}(t) & = & \frac{Z^{05}}{M_{PS}}\exp(-\frac{1}{2}M_{PS}T)
\sinh(M_{PS}(\frac{T}{2}-t))\; ,\nonumber\\
& & \nonumber\\
G_{n,\delta}(t) & = & C^{n,\delta}
\exp(-M_{n,\delta}t)\; ,\nonumber
\end{eqnarray}
in the time intervals reported in table~1. 
In eqs. \ref{funzfit}, $T$ represents the lattice time extension, the
subscripts $PS$ and $V$ stand for pseudoscalar and vector meson, $n$ and
$\delta$ stand for nucleon- and delta-like baryons\footnote{i.e. $n$ stands for
the nucleon $N$, $\Lambda\Sigma$ or $\Xi$ baryon, and $\delta$ is either
a $\Delta^{++}$ or a $\Omega$.}.                  
To improve stability, the meson (axial-pseudoscalar)
correlation functions have been symmetrized (anti-symmetrized) around $t=T/2$. 
The time fit intervals have been chosen with the following criteria: we fix the lower limit
of the interval as the one at which there is a stabilization of the effective 
mass, 
and, as the upper limit the furthest possible point before the error overwhelms
the signal.\\  
The pseudoscalar and vector decay constants $f_{PS}$ and $1/f_V$ are 
defined through the equations
\begin{eqnarray}
\langle 0|A_0|PS \rangle & = & i \frac{f_{PS}}{Z_A} M_{PS}\; ,\label{za}\\
\langle 0|V_i|V,r \rangle & = & \epsilon^r_i \frac{M_V^2}{f_V Z_V}\; ,
\end{eqnarray}
where $\epsilon^r_i$ is the vector-meson polarization, $M_{PS}$ and $M_V$ are
the pseudoscalar and vector masses and $Z_{V,A}$ are the renormalization
constants.
We have extracted $f_{PS}$ from the ratio
\begin{eqnarray}
R_{f_{PS}}(t) & = & Z_A \frac{G_{05}(t)}{G_{55}(t)}\nonumber\\
& \longrightarrow & Z_A \frac{Z^{05}}{Z^{55}}\tanh(M_{PS}(\frac{T}{2}-t))\nonumber\\  
& = & Z_A \frac{\langle 0|A_0|P \rangle}{\langle 0|P_5|P\rangle}
      \tanh(M_{PS}(\frac{T}{2}-t))\nonumber\\
& = & \frac{f_{PS} M_{PS}}{\sqrt{Z^{55}}}\tanh(M_{PS}(\frac{T}{2}-t))\; ,
\end{eqnarray}
and the vector-meson decay constant has been obtained directly from the 
parameters of the fit to $G_{ii}(t)$, eqs. \ref{funzfit}:
\begin{eqnarray}
\frac{1}{Z_V f_V} = \frac{\sqrt{Z^{ii}}}{M_V^2} \; .
\end{eqnarray}

\vspace{1mm}

\noindent
{\bf Light-Light hadron masses and decay constants}  \\

\noindent
In order to reduce the error coming from the chiral extrapolation, 
we have extracted as much physics as possible from the
``strange" region. The method we have used to extract the 
hadron masses and the light-light mesons decay constants from the lattice data 
is fully explained in \cite{mio1}.
\setlength{\tabcolsep}{.6pc}
\begin{table}
\caption{Predicted hadron masses in $GeV$ for all lattices, using  the scale 
from $M_{K^*}$.}
\label{tab:baryonsGeV}
\begin{center}
\begin{tabular}{||c|ccccccc||}
\hline\hline
& $M_\rho$ & $M_\phi$       
& $M_N $ & $M_{\Lambda\Sigma}$ & $M_\Xi $& $M_\Delta $ & $M_\Omega $\\
\hline
Exper.  & 0.770    & 1.019  & 0.9389  & 1.135   & 1.3181  & 1.232   & 1.6724 \\
\hline
C60a   & 0.809(7) & 0.977(7) 
& 1.09(5) & 1.21(4) & 1.32(4) & 1.3(1)  & 1.60(9) \\   
W60  & 0.808(3) & 0.978(3) 
& 1.19(5) & 1.29(4) & 1.40(4) & 1.46(7) & 1.71(4) \\   
C62a & 0.81(1)  & 0.98(1) 
& 1.1(1)  & 1.22(8) & 1.34(7) &  -      &  -   \\   
W62a  & 0.803(6) & 0.984(6) 
& 1.17(7) & 1.28(6) & 1.39(5) &  -      &  -   \\   
C62b   & 0.79(1)  & 1.00(1) 
& 1.1(2)  & 1.2(2)  & 1.4(1)  & 1.6(3)  & 1.9(2) \\   
W62b  & 0.797(7) & 0.989(7)
& 1.2(1)  & 1.3(1)  & 1.40(9) & 1.50(9) & 1.72(5) \\   
W64 & 0.796(4) & 0.990(4)
& 1.21(9) & 1.32(8) & 1.43(6) & 1.4(2)  & 1.72(9) \\   
C64& 0.792(4) & 0.994(4)
& 1.2(1)  & 1.29(8) & 1.41(7) & 1.3(2)  & 1.7(1) \\   
\hline
\hline
\end{tabular}
\end{center}
\end{table}
\setlength{\tabcolsep}{1.2pc}
\begin{table}
\caption{Predicted meson decay constants for all lattices, 
using  the scale from $M_{K^*}$.}
\label{tab:decaysGeV}
\begin{center}
\begin{tabular}{||c|ccccc||}
\hline\hline
&$f_\pi$~(GeV)& 
$\displaystyle\frac{1}{f_\rho}$&$f_K$~($GeV$) 
&$\displaystyle\frac{1}{f_{K^*}}$& $\displaystyle\frac{1}{f_{\phi}}$\\
\hline
Exper.  & 0.1307   & 0.28  & 0.1598   &   & 0.23\\
\hline
C60a   & 0.134(9) & 0.35(3) & 0.149(8) & 0.33(2) & 0.30(1)\\ 
W60  & 0.155(7) & 0.37(2) & 0.167(6) & 0.35(1) & 0.324(8)\\ 
C62a & 0.124(9) & 0.33(3) & 0.143(8) & 0.30(2) & 0.281(7)\\ 
W62a  & 0.135(6) & 0.35(2) & 0.153(5) & 0.33(1) & 0.307(6)\\ 
C62b   & 0.14(2)  & 0.26(4) & 0.16(2)  & 0.25(3) & 0.25(2)\\ 
W62b  & 0.135(8) & 0.36(2) & 0.152(7) & 0.34(2) & 0.315(9)\\ 
W64 & 0.147(9)  & 0.29(1) & 0.161(8) & 0.283(9) & 0.272(6)\\ 
C64& 0.144(9) & 0.25(1) & 0.158(8) & 0.25(1) & 0.243(8)\\ 
\hline
\hline
\end{tabular}
\end{center}
\end{table}
In tables~\ref{tab:baryonsGeV}~-~\ref{tab:decaysGeV}, we give a list of 
the lattice predictions
for the light mass spectrum and light-light decay constants from C60a~-~C64. The scale
is set from the $K^*$ mass.
Note that all errors quoted in tables~\ref{tab:baryonsGeV}~-~\ref{tab:decaysGeV} 
are statistical only.\\
As can be seen from table~\ref{tab:baryonsGeV}, for the hadron masses there is
good consistency of the physical predictions between the different simulations
within the errors.\\
To compare the lattice decay constants with the experimental ones, we have used
a `boosted' one-loop form of the renormalization constants
\cite{zwang}~-~\cite{gabrielli}:
\begin{eqnarray} 
\mbox{Wilson action}\; & Z_A & =  1 - 0.134 g_{\overline{MS}}^2  \nonumber\\ 
                       & Z_V & =  1 - 0.174 g_{\overline{MS}}^ 2  \nonumber\\ 
\mbox{Clover action}\; & Z_A & =  1 - 0.0177 g_{\overline{MS}}^2 \nonumber\\ 
                       & Z_V & =  1 - 0.10  g_{\overline{MS}}^2  \nonumber
\end{eqnarray}
where $g_{\overline{MS}}^2 = 6/\beta_{\overline{MS}}$,
with 
\begin{equation}
\beta_{\overline{MS}} = <U_{plaq}> \beta + 0.15.
\end{equation}
The results are shown in table~\ref{tab:decaysGeV}.
For the ratio $f_{PS}/{M_V}$ we see again a
global consistency  of our data both for Wilson and the SW-Clover
action. We do not see a strong $a$ dependence for either actions comparing
lattices at different $\beta$ (C60a, C62a and W60, W62a). The experimental 
points lie quite well in the
extrapolated/interpolated lattice data.\\
The values of the pseudoscalar decay constants in table~\ref{tab:decaysGeV}
show larger deviations from the experimental data than the ratio
$f_{PS}/{M_V}$. It is clear that a cancellation of systematic error occurs in
this ratio. Still there is an agreement within 1.5 standard deviations with the
experimental data apart from the W60 value which is quite high.\\ 
The situation is more delicate for the vector decay constant for which a
dependence on the volume and $a^{-1}$ may be present but with our data it 
would be difficult to disentangle the two effects \cite{mio1}. The
`strange' vector decay constants seem to be slightly more stable also because
no extrapolation is needed.
\\

\vspace{3mm}

\noindent
{\bf Strange Quark mass} \\

\noindent
Lattice QCD is in principle able to predict the mass of a quark from
the experimental value of the mass of a hadron containing that quark. The 
`bare' lattice quark mass $m(a)$ can be extracted directly from
lattice simulations and can be related to the continuum mass 
$m^{\overline{MS}}(\mu)$ renormalized in the minimal-subtraction dimensional
scheme through a well-defined perturbative procedure
\cite{qmass1}.  Following ref.~\cite{qmass1}
\[
m^{\overline{MS}}(\mu) = Z_m^{\overline{MS}}(\mu a)m(a)
\] 
where $m(a)$ is the bare lattice quark mass 
and $Z_m^{\overline{MS}}(\mu a)$ is the mass renormalization constant at scale
$\mu$ which we choose to be $2~GeV$. The bare and renormalized strange
quark masses are reported in table~\ref{tab:quarkm}.
\setlength{\tabcolsep}{1.2pc}
\begin{table}
\caption{Values for the lattice strange lattice quark masses for all 
lattices and the corresponding $\overline{MS}$ values at NLO, both in $MeV$. }
\label{tab:quarkm}
\begin{center}
\begin{tabular}{||c|ccc||}
\hline\hline
       & $m_s(a)$ & $m_s^{\overline{MS}}(\mu=2\;\mbox{$GeV$})$ & \\
\hline
C60a    &  89(3)  &  120(10)   &  \\
W60    &  98(2)  &  130(20)   &  \\
C62a   &  83(4)  &  120(10)   &  \\
W62a   &  93(3)  &  130(10)   &  \\
C62b   &  75(6)  &  110(10)   &  \\
W62b   &  92(4)  &  130(20)   &  \\
W64    &  82(3)  &  120(10)   &  \\
C64    &  69(3)  &  100(10)   &  \\
\hline
\hline
\end{tabular}
\end{center}
\end{table}
The error on $m^{\overline{MS}}(\mu)$ has been estimated as in ref.~\cite{qmass1}
taking into account the spread due to different definitions of the strong
coupling constant. 
There is a good consistency among the values coming from the different
lattices and we do not see any dependence on the lattice spacing $a$
within the errors in the Wilson data and a mild tendency in the clover data to
decrease with increasing $\beta$.
We therefore conclude that any $O(a)$ effects present are beneath the
level of statistics, and/or hidden among finite volume effects.
We then extract the average value of the
strange quark mass without any extrapolation and get:
\[
m_s^{\overline{MS}}(\mu = 2 GeV) = 122 \pm 20 MeV
\]
which is in agreement with the result of ref.~\cite{qmass1}. It is also
compatible with the value $m_s^{\overline{MS}}(\mu = 2 GeV) = 100 \pm 21 
\pm 10\; MeV$ of
ref.~\cite{guptamstrange}, but one should take into account that this value
comes from an analysis on various Wilson and Staggered lattices at different
values of $\beta$ and an extrapolation in $a$.
\\

\vspace{3mm}

\noindent
{\bf Heavy-Light decay constants}  \\

\noindent
The major sources of uncertainty in the determination of the heavy-light
pseudoscalar decay constants, 
besides the effects due to the use of the quenched approximation, come
from the calculation of the constant $Z_A$ in eq.~(\ref{za}) and
from  discretization errors of $O(a)$ present in the  operator matrix elements. 
A method to get rid of $Z_A$ consists of extracting 
the decay constants of  heavier pseudoscalar
mesons by computing the ratio $R_P = f_P / f_{\pi}$
and multiplying $R_P$ by the experimental value of the pion decay constant.
Hopefully, by taking the ratio, some of the  $O(a)$ effects are eliminated.
These  effects  are expected to be more important for $f_D$ than for 
$f_{\pi,K}$ since, at current values of $a$, the relevant 
parameter $m_{{\rm charm}} a $ is not very small. A full analysis of the
discretization errors is performed in the section 2 of \cite{mio2}.    
The results are
reported in table~\ref{tab:hevydecaysGeV}\footnote{For the run W60, the 
heavy-light correlators are computed on a subset of $120$ configurations.}.
We will first discuss the results for charmed mesons, for which the extrapolation in the heavy
quark mass is not a relevant source of systematic uncertainties, and
then discuss the $B$-meson case.
\setlength{\tabcolsep}{.6pc}
\begin{table}
\caption{Predicted heavy-light meson decay constants for all lattices.}
\label{tab:hevydecaysGeV}
\begin{center}
\begin{tabular}{||c|cccccc||}
\hline\hline
Run & $f_{D_s}/f_K$ & $f_{D}/f_\pi$ & $f_{D_s}/f_D$ & $f_{B_s}/f_K$ 
    & $f_{B}/f_\pi$ & $f_{B_s}/f_B$\\
\hline
C60b   & 1.56(3) & 1.63(4) & 1.08(1) & 1.48(7) & 1.53(9) & 1.10(3)\\       
C62a   & 1.48(6) & 1.58(8) & 1.07(2) & 1.28(9) & 1.32(13)& 1.14(6)\\     
W$60^3$& 1.11(3) & 1.14(4) & 1.06(1) & 0.79(4) & 0.81(5) & 1.05(2)\\       
W62a   & 1.23(4) & 1.31(6) & 1.07(1) & 0.83(4) & 0.86(6) & 1.10(3)\\     
W62b   & 1.19(5) & 1.25(7) & 1.09(2) & 0.84(5) & 0.86(6) & 1.12(3)\\ 
\hline
\hline
\end{tabular}
\end{center}
\end{table}
We find that higher statistics and larger intervals
of $a$ values are required to satisfactorily uncover the $a$
dependence of the decay constants. An extrapolation to $a=0$
it is not possible at this stage. Thus we believe that the best estimate
of $f_{D_s}$ is obtained from the Clover data at $\beta=6.2$, by using
$R_{D_s} = f_{D_s}/f_K$ (from a linear fit in the light quark masses,
a quadratic fit in $1/M_{P_s}$ and without any KLM factor).
As for the error, we take as a conservative
estimate of the  discretization error the difference between the results
 obtained
at $\beta=6.0$ and $6.2$, and combine it in quadrature with the statistical
one. This gives $R_{D_s}=1.48(10)$ from which, by using $f_K^{\exp}=
159.8$~MeV, we obtain $f_{D_s}=237 \pm 16$~MeV. This value is in very good 
agreement with the experimental value \cite{richman}\footnote{There is a new 
measurement of  
$B(D_s^{-}\rightarrow \tau^{-}\bar{\nu}_\tau)$ from L3 \cite{l3}. They have 
obtained
$B(D_s^{-}\rightarrow \tau^{-}\bar{\nu}_\tau) = (0.074 \pm 0.028(\mbox{stat}) 
\pm 0.016(\mbox{syst})\pm 0.018(\mbox{norm}))$ which corresponds to 
$f_{D_s} = (309 \pm 58(\mbox{stat}) 
\pm 33(\mbox{syst})\pm 38(\mbox{norm}) )$~MeV.}
of branching fraction 
$B(D_s^{+}\rightarrow \mu^{+}\nu_\mu) = (4.6 \pm 0.8 \pm 1.2)\times 10^{-3}$
which corresponds to a decay constant value of $f_{D_s} = (241 \pm 21 \pm
30)$ MeV (note that $f_{D_s}$ was predicted by lattice calculations long
before its measurement). By using $R_{D_s}=1.48(10)$
combined with $f_{D_s}/f_D=1.07(4)$ we obtain $f_D=221 \pm 17$~MeV.
We believe that these are our ``best" results.\\ 
In order to obtain $f_{B_s}$ and $f_B$, an extrapolation in the heavy quark
mass well outside the range available in our simulations is
necessary. Discretization errors can affect the final results in two ways.
Not only they can change the actual values of the decay constants, but also
deform the  dependence of $f_P$ on $m_H$. Moreover, points
obtained at the largest values of $m_H a$ become the most important, since
we extrapolate in the direction of larger values of $m_H$. A full
analysis of these effects is done in \cite{mio2}.\\
In order to extract our ``best" values
for $f_{B_s}$ and $f_{B}$ we have proceed exactly as for the $D$-meson case.  
We have obtained $f_{B_s}=205 \pm 15 \pm 31$~MeV=
$205 \pm 35$~MeV, where the second error ($31$~MeV) is the discretization
error, estimated by comparing the results from C60b and C62a as done for 
$f_{D_s}$.
We also have obtained $f_B = 180 \pm 32$~MeV. The decay constants
of the $B$-mesons are not yet measured and the numbers given above are
predictions of the lattice. The results we have obtained for the heavy-light
decay constants are in very good agreement with the compilations of lattice
calculations presented in refs.~\cite{flynn,beauty96}.   
\\

\vspace{3mm}

\noindent
{\bf Conclusions}  \\ 

\noindent
In this talk we have reported results from a large set of data on lattices of 
different lattice spacing and lattice sizes obtained with both the Wilson and 
SW-Clover actions. The results obtained and reported in the abstract
give us confidence in the ability of the lattice to predict other 
non-perturbative quantities that are of phenomenological interest. 
Further studies, with comparable (or smaller) statistical
errors and physical volume,
at smaller values of the lattice spacing, are required to reduce 
the systematic error due to O(a) effects.  
The systematic errors will be completely understood only when the quenched
approximation will be removed.
\\

\vspace{3mm}

\noindent
{\bf Acknowledgements}  \\

\noindent
The results presented here have been obtained in a enjoyable collaboration
with C.A. Allton, L. Conti, M. Crisafulli, V. Gim\'enez, G. Martinelli and 
F. Rapuano to whom I am much indebted. I would like to thank R. Barbieri for
his help in preparing my transparencies. Finally I warmly thank the 
organizers of the
XXV ITEP Winter School for organizing such a stimulating and enjoyable school.

\end{document}